\documentclass[12pt,english,eqsecnum,english,prd,aps,nofootinbib,superscriptaddress,amsfonts,longbibliography,tightenlines,showpacs]{revtex4}
\usepackage[T1]{fontenc}
\usepackage[utf8]{inputenc}
\usepackage{color}
\usepackage{babel}
\usepackage{mathtools}
\usepackage{amsmath}
\usepackage{amssymb}
\usepackage{stmaryrd}
\usepackage{cancel}
\usepackage{enumitem}
\usepackage[unicode=true,pdfusetitle,
 bookmarks=true,bookmarksnumbered=false,bookmarksopen=false,
 breaklinks=true,pdfborder={0 0 1},backref=false,colorlinks=false]
 {hyperref}

\usepackage{mathrsfs}


\begin{document}
%\title{Particle creation in loop quantum cosmology}
\title{Cosmological particle production in quantum gravity}

\author{Yaser Tavakoli}
\email{yaser.tavakoli@guilan.ac.ir}

\affiliation{Department of Physics, University of Guilan, 41335-1914 Rasht, Iran}
\affiliation{School of Astronomy, Institute for Research in Fundamental Sciences (IPM),	 19395-5531, Tehran, Iran}

\date{\today}
\begin{abstract}
Quantum theory of a  test field  on a  quantum cosmological spacetime may be viewed as a theory of the test field on  an emergent classical background. In such a case, the resulting dressed metric for the field propagation is a function of the quantum fluctuations of the original geometry. 
When the backreaction is negligible, massive modes can experience an anisotropic Bianchi type I background.  
The field modes propagating  on such a quantum-gravity-induced spacetime can then unveil  interesting phenomenological consequences of the super-Planckian scales, such as gravitational particle production.
The aim of this paper is to address the  issue of gravitational particle production associated to the massive modes in  such an anisotropic dressed spacetime. By imposing a suitable adiabatic condition on the vacuum state and computing the energy density of the created particles, the significance of the particle production on the dynamics of the  universe in Planck era  is discussed.
\end{abstract}

\maketitle

\section{Introduction}

Observation of cosmic microwave background (CMB) implies that the 
classical universe is homogeneous and isotropic on scales larger than 250 million light years. Based on the standard $\Lambda$CDM model of cosmology, the Friedmann-Lema\^itre-Robertson-Walker (FLRW) 
solution of Einstein's field equations,  
provides a suitable explanation of such near perfect
isotropy of the CMB and other astrophysical observables.
Nevertheless, there is not yet a decisive answer to the question whether or not the quantum structure of the universe  in the super-Planckian regime has the same symmetry as observed in the CMB, and how  such structure can be traced in the observational data.

According to the  standard model of cosmology, the structure formation at large scales are described  via (inhomogeneous) perturbations at smaller scales, in the early universe.
Quantum field theory in classical, curved spacetime provides a good approximate description of such phenomena in a regime
where the quantum effects of gravity are negligible \cite{Birrell:1982ix,Parker:2009uva}.
Within such framework, the issue of the gravitational particle production induced by  the time-dependent  background in an expanding universe and its backreaction effect  is of significant importance  \cite{Hu:1978zd,Zeldovich:1970si}.
Nevertheless, when tracing further back in time, where the  curvature of the universe reaches the Planck scales, the quantum effects of gravity becomes important. This might lead to additional phenomena which are expected to be important when exploring the dynamics of the early universe.

Loop quantum cosmology (LQC) is a promising candidate  to
investigate the quantum gravity effects in Planckian regime \cite{Ashtekar:2003hd}. 
It follows the quantization scheme of loop quantum gravity (LQG), which is a background independent, non-perturbative approach
to quantization of general relativity \cite{Ashtekar:2004eh,Rovelli:2004tv,Thiemann:2007zz}.
This approach has provided a number of concrete results: the classical big bang singularity is resolved and  is replaced by a quantum bounce \cite{Ashtekar:2008zu,Ashtekar:2006uz,Ashtekar:2009vc,Ashtekar:2011ni}; the standard theory of cosmological perturbations has been extended to 
a self-consistent theory from the bounce in the super-Planckian regime to the onset of slow-roll inflation \cite{Ashtekar:2009mb,Agullo:2012fc,Agullo:2013ai,Agullo:2012sh,Ashtekar:2015dja}.
Further phenomenological consequences  and observational predictions of LQC can be found e.g., in Refs.~\cite{Dapor:2012jg, Dapor:2012qn,Lewandowski:2017cvz,Tavakoli:2014mra,Graef:2020qwe,Garcia-Chung:2020zyq,Tavakoli:2015fvz,Lankinen:2016ile,Scardua:2018omf,Assaniousssi:2014ota,Celani:2016cwm} and \cite{Ashtekar:2015dja,Bolliet:2015raa}, respectively.

The cosmological, quantum backgrounds in LQC establish a fruitful ground to 
explore the quantum theory of inhomogeneous  fields  propagating on it. In a dressed metric approach, when the full quantum Hamiltonian constraint of the gravity-matter system is solved, an effective dressed geometry emerges for the field modes. Depending on whether the field is massless or massive, various spacetime metrics can raise for the dressed background geometry. The emergent metric components, if  the backreaction between the fields and the geometry is discarded, depend, generically, on the  fluctuations of the original quantum metric operator only. For massive modes, quantum gravity effects may induce a small deviation from the initial isotropy  \cite{Dapor:2012qn,Tavakoli:2015fvz}. If the backreaction is considered, other properties, such as emerging a mode-dependent  background, can raise, which lead to the violation of the local Lorentz symmetry \cite{Dapor:2012jg,Lewandowski:2017cvz}. Even in the absence of the backreaction between the states of the gravity and matter fields, a backreaction effect may arise due to the gravitational particle production  which can challenge the validity of the test field approximations employed in the dressed metric scenario \cite{Graef:2020qwe}.

Motivated from the above paragraphs, our purpose in this article is to address some  phenomenological issues associated to the massive test field propagating on a quantum cosmological spacetime. In particular, we will consider a scenario in which an anisotropic dressed geometry emerges for the massive modes due to quantum gravity effects, and then 
revisit the occurrence of the  gravitational particle production  on such an  anisotropic dressed background. Such effects will have important consequences in the   super-Planckian regime and at the onset of  the classical inflationary epoch. 
We thus organize this paper as follows. 
In Sec.~\ref{QFT-FLRW}, we will consider the propagation of a massive test field on a  quantized FLRW spacetime. We will derive an effective evolution equation for the field and obtain a suitable anisotropic dressed  background for the field propagation.
In Sec.~\ref{fluctuation}, we will study the theory of quantum field,  by considering the infinitely many modes, on the emergent anisotropic spacetime. We will then explore the problem of the gravitational particle production by choosing a convenient, adiabatic vacuum state in the super-Planckian regime. 
Finally, in Sec.~\ref{conclusion} we will
present the conclusion and discussion  of our work.

\section{Quantum Fields on a quantum FLRW spacetime}
\label{QFT-FLRW}

In this section, we study the quantum  theory of a massive scalar field  on a quantum FLRW background. Firstly, we will analyze the quantum theory of a scalar field in the classical Friedmann universe. Next, we will quantize the background and obtain a quantum evolution equation for the composite state of the gravity-perturbation system. Finally, in the last step, we will extract an effective evolution equation for the field mode degrees of freedom due to which we can extract an effective {\em anisotropic} background for propagation of the massive modes.

\subsection{Quantum field on a  cosmological classical spacetime}
\label{ClassicalBack}

We start 
with considering  a {\em massive} scalar field propagating on a more general classical, \emph{anisotropic} background spacetime. In particular, we take a Bianchi type I model whose line element is  
 represented by 
\begin{equation}
g_{ab}dx^{a}dx^{b}=-N_{x_{0}}^{2}\left(x_{0}\right)\left(dx_{0}\right)^{2}+\sum_{i}^{3}a_{i}^{2}\left(x_{0}\right)\left(dx^{i}\right)^{2},
\label{BI-metric-class}
\end{equation}
where, $N_{x_{0}}$ is the lapse function and $a_{i}$ the scale
factor in the $x^i$-direction. The coordinates
$(x_{0},\mathbf{x})$ are chosen such that, $x_{0}\in\mathbb{R}$
is a generic time coordinate, and $\mathbf{x}\in\mathbb{T}^{3}$ (i.e.,  a 3-torus
with the coordinates $x^{j}\in(0,\ell_{j})$).

We consider a real, minimally coupled scalar field, $\varphi(x_{0},\mathbf{x})$,
with a mass $m$, propagating on the background (\ref{BI-metric-class}). 
Having the Lagrangian density for $\varphi(x_{0},\mathbf{x})$, with a quadratic potential,
\begin{equation}
\mathcal{L}_\varphi = -\frac{1}{2}\left[g^{ab}\nabla_a\varphi\nabla_b\varphi + m^2\varphi^2\right],
\end{equation}
the equation of motion  is obtained as
\begin{equation}
(g^{ab}\nabla_a\nabla_b-m^{2})\varphi=0.
\label{KG-equation}
\end{equation}
On the $x_0 = const$ slice, by performing the Legendre transformation, we take a
canonically conjugate momentum $\pi_\varphi$ to the field $\varphi$. Then,  the classical solutions of Eq.~(\ref{KG-equation}) for the pair $(\varphi, \pi_\varphi)$ can be Fourier expanded as\footnote{We  consider an
	elementary cell $\mathcal{V}$ by fixing its edge to lie along the coordinates $(\ell_1, \ell_2, \ell_3)$. We then denote  the volume of $\mathcal{V}$ by $\mathring{V}=\ell_1\ell_2\ell_3\equiv \ell^3$, and  restrict all integrations in the Fourier integral to this volume.}
\begin{subequations}
	\begin{equation}
	\varphi(x_{0},\mathbf{x})=\frac{1}{\ell^{3/2}}\sum_{\mathbf{k}\in\mathscr{L}}\varphi_{\mathbf{k}}(x_{0})\, e^{i\mathbf{k}\cdot\mathbf{x}},
	\label{total-field1a}
	\end{equation}
	\begin{equation}
	\pi_{\varphi}(x_{0},\mathbf{x})=\frac{1}{\ell^{3/2}}
	\sum_{\mathbf{k}
		\in \mathscr{L}}\pi_{\mathbf{k}}(x_{0})\, e^{i\mathbf{k}\cdot\mathbf{x}}, \label{total-field1-a}
	\end{equation}
\end{subequations}
with  $\mathscr{L}=\mathscr{L}_{+}\cup\mathscr{L}_{-}$ being a 3-dimensional
lattice: 
\begin{subequations}
	\begin{eqnarray}
	\mathscr{L}_+ &=&  \{\mathbf{k}:~ k_3>0\} \cup \{\mathbf{k}:~ k_3=0, k_2>0\}  \cup \{\mathbf{k}: ~k_3=k_2=0, k_1>0\} , \label{lattice1}
	\end{eqnarray}
	and
	\begin{eqnarray}
	\mathscr{L}_- &=&  \{\mathbf{k}: ~k_3<0\} \cup \{\mathbf{k}:~ k_3=0, k_2<0\}   \cup \{\mathbf{k}:~ k_3=k_2=0, k_1<0\}, \label{lattice2}
	\end{eqnarray}
\end{subequations}
spanned by $\mathbf{k}=(k_{1},k_{2},k_{3})\in(2\pi\mathbb{Z}/\ell)^{3}$, where,
$\mathbb{Z}$ is the set of integers \cite{Ashtekar:2009mb,Dapor:2012jg}.

Let us decompose the field modes into the real and imaginary parts as
\begin{equation}
\varphi_{\mathbf{k}}(x_{0})=\frac{1}{\sqrt{2}}\left[\varphi_{\mathbf{k}}^{(1)}(x_{0})+i\varphi_{\mathbf{k}}^{(2)}(x_{0})\right].\label{decomposition-field}
\end{equation}
Then, reality condition  implies that 
%\begin{equation}
$\varphi_{\mathbf{k}}^{(1)}=\varphi_{-\mathbf{k}}^{(1)}$ and $\varphi_{-\mathbf{k}}^{(2)}=-\varphi_{\mathbf{k}}^{(2)}$.
%\end{equation}
By  introducing a new variable $Q_{\mathbf{k}}$, 
\begin{equation}
Q_{\mathbf{k}}(x_{0})=\begin{cases}
\varphi_{\mathbf{k}}^{(1)}(x_{0}), & \text{if \ensuremath{\mathbf{k}\in\mathscr{L}_{+}}},\vspace{2mm} \\
\varphi_{-\mathbf{k}}^{(2)}(x_{0}), & \text{if \ensuremath{\mathbf{k}\in\mathscr{L}_{-}}},
\end{cases}\label{q-variable}
\end{equation}
associated
to the real variables $\varphi_{\mathbf{k}}^{(1)}$ and $\varphi_{\mathbf{k}}^{(2)}$,
we can rewrite  the field modes $\varphi_{\mathbf{k}}$ as 
\begin{equation}
\varphi_{\mathbf{k}}(x_{0})=\frac{1}{\sqrt{2}}\left[Q_{\mathbf{k}}(x_{0})+iQ_{-\mathbf{k}}(x_{0})\right].\label{decomposition-field2}
\end{equation}
Likewise, decomposition of the momentum $\pi_{\mathbf{k}}$ as
\begin{equation}
\pi_{\mathbf{k}}(x_{0})=\frac{1}{\sqrt{2}}\left[\pi_{\mathbf{k}}^{(1)}(x_{0})+i\pi_{\mathbf{k}}^{(2)}(x_{0})\right],
\label{decomposition-momentum}
\end{equation}
in terms of the real variables $\pi_{\mathbf{k}}^{(1)}$ and $\pi_{\mathbf{k}}^{(2)}$, and introducing the new  variable $P_{\mathbf{k}}$ (conjugate
to $Q_{\mathbf{k}}$ above), as
\begin{equation}
P_{\mathbf{k}}(x_{0})=\begin{cases}
\pi_{\mathbf{k}}^{(1)}(x_{0}), & \text{if \ensuremath{\mathbf{k}\in\mathscr{L}_{+}}},\vspace{2mm}\\
\pi_{-\mathbf{k}}^{(2)}(x_{0}), & \text{if \ensuremath{\mathbf{k}\in\mathscr{L}_{-}}}\,,
\end{cases}\label{p-variable}
\end{equation}
yields,
\begin{equation}
\pi_{\mathbf{k}}(x_{0})=\frac{1}{\sqrt{2}}\left[P_{\mathbf{k}}(x_{0})+iP_{-\mathbf{k}}(x_{0})\right].
\label{decomposition-momentum2}
\end{equation}
The conjugate variables $(Q_{\mathbf{k}}, P_{\mathbf{k}})$
satisfy the Poisson bracket, $\{Q_{\mathbf{k}},P_{\mathbf{k}^{\prime}}\}=\delta_{\mathbf{k},\mathbf{k}^{\prime}}$.

Using  the variables (\ref{q-variable}) and (\ref{p-variable}), the Hamiltonian of the scalar test field, $\varphi$, can be written as the sum of the Hamiltonians,
$H_{\mathbf{k}}\left(x_{0}\right)$, of the decoupled harmonic oscillators,
each given in terms of $(Q_{\mathbf{k}},P_{\mathbf{k}})$:
\begin{equation}
H_{\varphi}(x_{0})\coloneqq\sum_{\mathbf{k}\in\mathscr{L}}H_{\mathbf{k}}(x_{0})=\frac{N_{x_{0}}(x_{0})}{2|a_{1}a_{2}a_{3}|}\sum_{\mathbf{k}\in\mathscr{L}}\Big(P_{\mathbf{k}}^{2}+\omega_{k}^{2}(x_{0})Q_{\mathbf{k}}^{2}\Big),\label{Hamiltonian-SF10}
\end{equation}
where, $\omega_{k}\left(x_{0}\right)$ is a time-dependent frequency,
defined by 
\begin{equation}
\omega_{k}^{2}\left(x_{0}\right)\coloneqq|a_{1}a_{2}a_{3}|^{2}\left[\sum_{i=1}^{3}\left(\frac{k_{i}}{a_{i}}\right)^{2}+m^{2}\right].\label{frequency-SF-b1}
\end{equation}
Note that, $Q_{\mathbf{k}}$ is the field amplitude
for the mode characterized by $\mathbf{k}$.
%, satisfying Eq.~(\ref{KG-equation}), which is the equation of motion
%obtained from  the Hamiltonian \eqref{Hamiltonian-SF10}.

To quantize the field, we follow  the Schr\"odinger representation.
While the background spacetime is left as classical,  quantization of $Q_\mathbf{k}$, for a fixed mode $\mathbf{k}$, resembles that of a quantum harmonic
oscillator with the Hilbert space {\footnotesize{}${\cal H}_{Q}^{(\mathbf{k})}=L^{2}(\mathbb{R},dQ_{\mathbf{k}})$}, where
$(Q_{\mathbf{k}}, P_{\mathbf{k}})$ are promoted to operators on  {\footnotesize{}${\cal H}_{Q}^{(\mathbf{k})}$} as
\begin{eqnarray}
\hat{Q}_{\mathbf{k}}\psi(Q_{\mathbf{k}})=Q_{\mathbf{k}}\psi(Q_{\mathbf{k}}) \quad \text{and} \quad \hat{P}_{\mathbf{k}}\psi(Q_{\mathbf{k}})=-i\hbar\big(\partial/\partial 
Q_{\mathbf{k}}\big)\psi(Q_{\mathbf{k}}).
\end{eqnarray}
Then,  the Hamiltonian operator $\hat{H}_{\mathbf{k}}$ generates
the time evolution of the state $\psi(Q_{\mathbf{k}})$ via
the Schr\"odinger equation 
\begin{equation}
i\hbar\partial_{x_{0}}\psi(x_{0},Q_{\mathbf{k}})=\frac{\ell^{3}N_{x_{0}}}{2V}\Big(\hat{P}_{\mathbf{k}}^{2}+\omega_{k}^{2}\hat{Q}_{\mathbf{k}}^{2}\Big)\psi(x_{0},Q_{\mathbf{k}}),\label{Hamiltonian-SF-bquantum1}
\end{equation}
where, $V=\ell^{3}\left|a_{1}a_{2}a_{3}\right|$ denotes the physical volume of the universe.

By setting $x_{0}=\phi$ in Eq.~\eqref{Hamiltonian-SF-bquantum1}, as an {\em internal time} parameter,  the evolution of the state $\psi(Q_{\mathbf{k}})$, with respect to  $\phi$,  on a  Bianchi-I background with   components $\big(\tilde{N}_\phi, \tilde{a}_i(\phi)\big)$,  reads
\begin{equation}
i\hbar\partial_{\phi}\psi\left(\phi,Q_{\mathbf{k}}\right)=\frac{\tilde{N}_{\phi}}{2\left|\tilde{a}_{1}\tilde{a}_{2}\tilde{a}_{3}\right|}\Big[\hat{P}_{\mathbf{k}}^{2}+\tilde{\omega}_k^2(\phi)\hat{Q}_{\mathbf{k}}^{2}\Big]\psi\left(\phi,Q_{\mathbf{k}}\right), \label{Hamiltonian-SF-bquantum-eff}
\end{equation}
where,
\begin{equation}
\tilde{\omega}_k^2(\phi)\, =\, \left(\sum_{i}^{3}\frac{\tilde{k}_{i}^2}{\tilde{a}_{i}^2}+\tilde{m}^{2}\right)\left(\tilde{a}_{1}\tilde{a}_{2}\tilde{a}_{3}\right)^{2}.
\label{Hamiltonian-SF-bquantum-eff0}
\end{equation}
Clearly, one gets an isotropic background for the field, by  $\tilde{a}_{1}(\phi)=\tilde{a}_{2}(\phi)=\tilde{a}_{3}(\phi)\equiv \tilde{a}(\phi)$.

\subsection{Quantization of the background \label{QuantumBack}}

In our model herein this paper, we will assume that the field,  $\varphi$, propagates on an {\em isotropic FLRW spacetime} in a super-Planckian regime, so that this isotropic background has to be quantized. However, the reason of constructing a general formalism of the field on an anisotropic background in the previous subsection, is for the purpose of comparison, when an effective dressed spacetime emerges from the isotropic quantum background. We will show that, the emergent effective spacetime can have the same structure of an anisotropic Bianchi-I geometry for the field propagation.

Let us assume a harmonic time gauge, $x_{0}=\tau$, and set the isotropic components, $a_1=a_2=a_3=a(\tau)$, in Eq.~(\ref{BI-metric-class}). Then, $N_{\tau}=a^{3}(\tau)$, 
%which relates to $N_{\phi}$  via $N_{\phi}=  %\left(\ell^{3}/p_{\phi}\right)N_{\tau}$.
%%
and  the Hamiltonian
\eqref{Hamiltonian-SF10}  becomes
\begin{equation}
H_{\varphi}^{{\rm (iso)}}=\sum_{\mathbf{k}}H_{\tau,\mathbf{k}}\coloneqq\frac{1}{2}\sum_{\mathbf{k}}\Big(P_{\mathbf{k}}^{2}+\omega_{\tau,k}^{2}\, Q_{\mathbf{k}}^{2}\Big),\label{Hamiltonian-SF-FLRW}
\end{equation}
where,
\begin{equation}
\omega_{\tau,k}^{2}(\tau) = k^{2}a^{4}(\tau) + m^{2} a^{6}(\tau).
\end{equation}
We note that,  the massless scalar field, $\phi(\tau)$, still serves as an internal time parameter.

In quantum theory, we will quantize not only the test field, but also the background geometry. We will assume that  the backreaction of the quantum field on the  background quantized spacetime is negligible. 
This yields an evolution for the wave function,  $\psi(Q_{\mathbf{k}})$, of the test field, with respect to the internal time $\phi$.
Let  ${\cal H}_{{\rm kin}}^{o}={\cal H}_{{\rm grav}}\otimes{\cal H}_{\phi}$ denotes the background Hilbert space, which consists of the Hilbert space of the gravity sector and that of the scalar clock variable $\phi$; the matter sector is quantized according to the Schr\"odinger representation, {\small{}${\cal H}_{\phi}=L^{2}(\mathbb{R},d\phi)$}.  
Likewise,  the Hilbert space of the  field modes reads
{\small{}${\cal H}_{\varphi}^{(\mathbf{k})}=L^{2}(\mathbb{R},dQ_{\mathbf{k}})$}, as before.
Subsequently, the full kinematical Hilbert space of the system, for a single mode $\mathbf{k}$, is given by {\small{}${\cal H}_{{\rm kin}}^{(\mathbf{k})}={\cal H}_{{\rm kin}}^{o}\otimes{\cal H}_{\varphi}^{(\mathbf{k})}$}.

In LQC, the background Hamiltonian  constraint operator,
\begin{eqnarray}
\hat{{\cal C}}_{o}=\hat{{\cal C}}_{{\rm grav}}+\hat{{\cal C}}_{\phi},
\end{eqnarray}
is well-defined on ${\cal H}_{{\rm kin}}^{o}$;
the physical states $\Psi_{o}(\phi,\nu)\in{\cal H}_{{\rm kin}}^{o}$
are those lying on the kernel of $\hat{{\cal C}}_{o}$, and are solutions
to the self-adjoint  constraint equation  \cite{Ashtekar:2006rx} 
\begin{equation}
N_{\tau}\hat{{\cal C}}_{o}\Psi_{o}(\nu,\phi)=-\frac{\hbar^{2}}{2\ell^{3}}(\partial_{\phi}^{2}+\Theta)\Psi_{o}(\nu,\phi)=0.\label{H-constraint1}
\end{equation}
The quantum number $\nu$ is the eigenvalue of the background volume operator, {\small{}$\hat{V}_{o}=\widehat{\ell^{3}a^{3}}$}, which acts on the states {\small{}$\Psi_{o}(\phi,\nu)\in{\cal H}_{{\rm kin}}^{o}$} as
\begin{equation}
\hat{V}_{o}\, \Psi_{o}(\nu,\phi)=2\pi\gamma\ell_{{\rm Pl}}|\nu|\, \Psi_{o}(\nu,\phi).
\end{equation}
Moreover, $\Theta$ is a difference operator that acts on $\Psi_{o}(\nu)$,
involving only the volume sector $\nu$.
Taking only the positive frequency solutions
to Eq. \eqref{H-constraint1}, we get  a Schr\"odinger equation
for the background as 
\begin{equation}
-i\hbar\partial_{\phi}\Psi_{o}(\nu,\phi)=\hbar\sqrt{\Theta}\Psi_{o}(\nu,\phi)\eqqcolon\hat{H}_{o}\Psi_{o}(\nu,\phi).
\label{geometry-evol}
\end{equation}
The  solutions yield a physical Hilbert space, ${\cal H}_{{\rm phys}}^{o}$, equipped by the inner product 
\begin{equation}
\langle\Psi_{o}|\Psi_{o}^{\prime}\rangle=\sum_{\nu}\Psi_{o}^{\ast}(\nu,\phi_{0})\Psi_{o}^{\prime}(\nu,\phi_{0}),
\end{equation}
for an ``instant'' $\phi_{0}$, of the internal time.

For a composite state  {\small{}$\Psi(\nu,Q_{\mathbf{k}},\phi)\in{\cal H}_{{\rm kin}}^{(\mathbf{k})}$} of the geometry-test field system, the action of the total quantum  Hamiltonian constraint, $\hat{{\cal C}}_{\tau,\mathbf{k}}$, is written as
\cite{Ashtekar:2009mb}
\begin{equation}
\hat{{\cal C}}_{\tau,\mathbf{k}}\Psi(\nu,Q_{\mathbf{k}},\phi)=\big(N_{\tau}\hat{{\cal C}}_{o}+\hat{H}_{\tau,\mathbf{k}}\big)\Psi(\nu,Q_{\mathbf{k}},\phi)=0,\label{const-tot}
\end{equation}
where,  $\hat{H}_{\tau,\mathbf{k}}$ is the Hamiltonian operator of the $\mathbf{k}$th field mode,
\begin{equation}
\hat{H}_{\tau,\mathbf{k}}=\frac{1}{2}\left[\hat{P}_{\mathbf{k}}^{2}+\left(k^{2}\hat{a}^{4}+m^{2}\hat{a}^{6}\right)\hat{Q}_{\mathbf{k}}^{2}\right].
\label{eq:H-tau-k}
\end{equation}
By replacing the expression \eqref{H-constraint1} into the constraint equation (\ref{const-tot}), we obtain
\begin{eqnarray}
-i\hbar\partial_{\phi}\Psi(\nu, Q_{\mathbf{k}}, \phi) =\left[\hat{H}_o^2 - 2\ell^{3}\hat{H}_{\tau,\mathbf{k}}\right]^{\frac{1}{2}}\Psi(\nu, Q_{\mathbf{k}}, \phi),
\label{const-tot1}
\end{eqnarray}
which represents  a quantum evolution of  $\Psi(\nu,Q_{\mathbf{k}},\phi)$
with respect to the internal  time $\phi$.
In a test field approximation, when the backreaction effect is omitted, the expression under the square root can be expanded, up to the first order terms, as \cite{Ashtekar:2009mb} 
\begin{equation}
-i\hbar\partial_{\phi}\Psi(\nu,Q_{\mathbf{k}},\phi)\, \approx\, \big(\hat{H}_{o}-\hat{H}_{\phi,\mathbf{k}}\big)\Psi(\nu,Q_{\mathbf{k}},\phi).
\label{shro-eq2-approx}
\end{equation}
In expanding the right-hand-side of Eq.~(\ref{const-tot1}) to derive the equation above, we  regarded $\hat{H}_{o}$ as the
Hamiltonian of the heavy degree of freedom, whereas $\hat{H}_{\phi,\mathbf{k}}$, defined by
\begin{equation}
\hat{H}_{\phi,\mathbf{k}}\, \coloneqq\, \ell^{3}\hat{H}_{o}^{-\frac{1}{2}}\hat{H}_{\tau,\mathbf{k}}\hat{H}_{o}^{-\frac{1}{2}},
\end{equation}
was considered as the Hamiltonian of the light degree of freedom (i.e, a perturbation term).
In this approximation, it is suitable to separate the total state of the system as
\begin{equation}
\Psi(\nu,Q_{\mathbf{k}},\phi)=\Psi_{o}(\nu,\phi)\otimes\psi(Q_{\mathbf{k}},\phi).
\end{equation}

To explore the quantum evolution of a pure 
test field state, $\psi(Q_\mathbf{k}, \phi)$,  on a time-dependent background, it is more convenient to employ an interaction picture. Thus, we introduce
 \begin{equation}
\Psi_{{\rm int}}(\nu,Q_{\mathbf{k}},\phi)=e^{-(i\hat{H}_{o}/\hbar)(\phi-\phi_{0})}\Psi(\nu,Q_{\mathbf{k}},\phi),
\label{interaction-pic}
 \end{equation}
In this picture, the geometry evolves by  $\hat{H}_{o}$,
through Eq.~(\ref{geometry-evol}), for any $\Psi_{o}\in{\cal H}_{{\rm kin}}^{o}$ in the Heisenberg picture, 
\begin{equation}
\Psi_{o}(\nu,\phi)=e^{(i\hat{H}_{o}/\hbar)(\phi-\phi_{0})}\Psi_{o}(\nu,\phi_{0}).
\end{equation}
Plugging this into Eq.~(\ref{interaction-pic}), we get
\begin{equation}
\Psi_{{\rm int}}(\nu,Q_{\mathbf{k}},\phi)=\Psi_{o}(\nu,\phi_{0})\otimes\psi(Q_{\mathbf{k}},\phi).
\label{interaction-pic2}
\end{equation}
Thereby, the geometrical sector in the composite state  $\Psi_{{\rm int}}(\nu,Q_{\mathbf{k}},\phi)$ becomes frozen in the instant of  time $\phi_0$, so that $\Psi_{{\rm int}}$ represents a time-dependent test field state, $\psi$, solely.

By replacing (\ref{interaction-pic2}) into the evolution equation \eqref{shro-eq2-approx}, and
tracing out the geometrical state, $\Psi_{o}(\nu,\phi_{0})$, a quantum evolution for $\psi(Q_{\mathbf{k}},\phi)$ is obtained  as
\begin{align}
i\hbar\partial_{\phi}\psi= & \frac{1}{2}\left[\langle\hat{H}_{o}^{-1}\rangle\hat{P}_{\mathbf{k}}^{2}+\left(k^{2}\big\langle \hat{H}_{o}^{-\frac{1}{2}}\hat{a}^{4}(\phi)\hat{H}_{o}^{-\frac{1}{2}}\big\rangle +m^{2}\big\langle \hat{H}_{o}^{-\frac{1}{2}}\hat{a}^{6}(\phi)\hat{H}_{o}^{-\frac{1}{2}}\big\rangle \right)\hat{Q}_{\mathbf{k}}^{2}\right]\psi,
\label{evol-eq1}
\end{align}
where, $\left\langle \cdot\right\rangle $ denotes the expectation value
with respect to  $\Psi_{o}(\nu,\phi_{0})$. 
It is clear that, the use of the interaction picture in Eq.~(\ref{evol-eq1})
provided a the Heisenberg description for the quantum geometrical elements; that is, the geometry state, $\Psi_{o}(\nu,\phi_{0})$, is fixed at time
$\phi=\phi_{0}$, while the geometrical operator, $\hat{a}(\phi)=\hat{V}_{o}^{1/3}(\phi)/\ell$,
 evolves in time as 
\begin{equation}
\hat{a}(\phi)=e^{-(i\hat{H}_{o}/\hbar)(\phi-\phi_{0})}\, \hat{a}\, e^{(i\hat{H}_{o}/\hbar)(\phi-\phi_{0})}.
\end{equation}
Therefore, Eq.~(\ref{evol-eq1}) represents a $\phi$-evolution of the field, $\psi(Q_{\mathbf{k}},\phi)$,
on a  {\em time-dependent} (classical) background, similar to the one we had in classical spacetime (cf. Eq.~(\ref{Hamiltonian-SF-bquantum-eff})).

\subsection{Emergence of anisotropic dressed spacetimes \label{DressedBack}}

Eq.~\eqref{evol-eq1} can be interpreted as an evolution equation for the field modes on a  (effective) classical spacetime, whose components are generated by the expectation values of the original isotropic quantum  geometry operators with respect to the unperturbed state $\Psi_o$. To explore the properties of such effective spacetime, we can compare Eq.~(\ref{evol-eq1}), for evolution of the state $\psi$, with the corresponding equation \eqref{Hamiltonian-SF-bquantum-eff}, for the same state $\psi$, on an anisotropic classical background.
This comparison yields a set of relations
between parameters of the  Bianchi I geometry, $(\tilde{N}_{\phi}, \tilde{a}_i, \tilde{k}_i, \tilde{m})$,  and those of the isotropic quantum geometry, $(\hat{a}, \hat{H}_o, k, m)$, as
\begin{subequations}
	\label{eq:main-sys}
	\begin{eqnarray}
	\tilde{N}_{\phi} &= & \ell^{3}\left|\tilde{a}_{1}\tilde{a}_{2}\tilde{a}_{3}\right|\big\langle \hat{H}_{o}^{-1}\big\rangle ,\label{eq:main-sys-1}\\
	\sum_{i=1}^{3}\frac{\tilde{k}_{i}^2}{\tilde{a}_{i}^2}\, \tilde{N}_{\phi}\, \left|\tilde{a}_{1}\tilde{a}_{2}\tilde{a}_{3}\right| &= & \sum_{i=1}^{3}k_{i}^{2}\ell^{3}\big\langle \hat{H}_{o}^{-1/2}\hat{a}^{4}\hat{H}_{o}^{-1/2}\big\rangle ,\label{eq:main-sys-2}\\
	\tilde{N}_{\phi}\, \tilde{m}^{2}\, \left|\tilde{a}_{1}\tilde{a}_{2}\tilde{a}_{3}\right| &= & \ell^{3}m^{2}\big\langle \hat{H}_{o}^{-1/2}\hat{a}^{6}\hat{H}_{o}^{-1/2}\big\rangle .\label{eq:main-sys-3}
	\end{eqnarray}
\end{subequations}
Note that, $\hat{a}\equiv\hat{a}(\phi)$ and $\langle\hat{H}_{o}^{-1}\rangle=(\tilde{p}_{\phi})^{-1}$.
Eqs.~(\ref{eq:main-sys}) provide an underdetermined system  of five equations 
 with eight unknowns $(\tilde{N}_{\phi}, \tilde{a}_i, \tilde{k}_i, \tilde{m})$. Thus, to be able to solve this system, we need
to impose some arbitrary conditions on these parameters to reduce the number of unknowns to five. 
Different classes of solutions, by imposing various conditions on the variables, and their physical consequences were discussed in \cite{Tavakoli:2015fvz}. 
As an example, we will present two classes of such solutions; one is produced by a massless test field, $m=\tilde{m}=0$, and the other is  provided by the {\em dressed} massive modes, $\tilde{m}, m\neq 0$.

For a massless test field, $m=0$, we will immediately obtain $\tilde{m}=0$. In this case, we will have four equations for the seven unknown parameters $(\tilde{N}_{\phi}, \tilde{a}_i, \tilde{k}_i)$. Therefore, we will still need three more conditions to be able
to solve the system (\ref{eq:main-sys-1}) and (\ref{eq:main-sys-2}). The simplest choice is  $k_{i}^{2}=\tilde{k}_{i}^{2}$
(for each $i$), so that  $\tilde{a}_{1}^{2}=\tilde{a}_{2}^{2}=\tilde{a}_{3}^{2}=\tilde{a}^{2}$. Then, we
obtain
\begin{subequations}
	\label{massless-dressed-scale}
\begin{eqnarray}
\tilde{a}^{4} &= & \frac{\big\langle \hat{H}_{o}^{-1/2}\hat{a}^{4}(\phi)\hat{H}_{o}^{-1/2}\big\rangle }{\big\langle \hat{H}_{o}^{-1}\big\rangle },\label{massless-dressed-scale-1}  \vspace{1mm}\\
\tilde{N}_{\phi} &= & \ell^{3}\big\langle \hat{H}_{o}^{-1}\big\rangle ^{\frac{1}{4}}\big\langle \hat{H}_{o}^{-1/2}\hat{a}^{4}(\phi)\hat{H}_{o}^{-1/2}\big\rangle ^{\frac{3}{4}}=\ell^{3}\tilde{p}_{\phi}^{-1}\tilde{a}^{3}\, \equiv\, \bar{N}_{\phi}.\label{massless-dressed-scale-2}
\end{eqnarray}
\end{subequations}

If $\,m,\tilde{m}\protect\neq0$ and $\tilde{m}\protect\neq m$, we will have  different ranges of solutions [cf. \cite{Tavakoli:2015fvz}]. However, for our purpose in this paper, we will  consider only a specific solution by imposing the condition $\tilde{k}_{i}=\alpha_{ii}k_{i}$ (with $i=1,2,3$).
This  yields
\begin{subequations}
	\label{anisotropyBG}
	\begin{eqnarray}
	\tilde{N}_{\phi} &= & \frac{\bar{N}_{\phi}}{\lambda}\, ,\label{lapse-massdressed1} \\
	\tilde{a}_{i} &= & \frac{\alpha_{ii}}{\lambda}\, \tilde{a}, \label{anisotropy-scale1}\\
	\tilde{m}^{4} &= & m^{4}\lambda^{4}\, \frac{\big\langle \hat{H}_{o}^{-1/2}\, \hat{a}^{6}(\phi)\, \hat{H}_{o}^{-1/2}\big\rangle ^{2}\big\langle \hat{H}_{o}^{-1}\big\rangle }{\big\langle \hat{H}_{o}^{-1/2}\hat{a}^{4}(\phi)\hat{H}_{o}^{-1/2}\big\rangle^{3}},
	\label{anisotropy-scale3}
	\end{eqnarray}
\end{subequations}
where, $\lambda\equiv (\alpha_{11}\alpha_{22}\alpha_{33})^{1/2}$ is a constant.
Note that, as a special subcase, when $\alpha_{ii}=1$ (so $\lambda=1$), 
the  wave vector becomes undressed, $\tilde{k}_{i}=k_{i}$, and an isotropic dressed scale factor, $\tilde{a}_{1}=\tilde{a}_{2}=\tilde{a}_{3}=\tilde{a}$,
identical to Eq.~\eqref{massless-dressed-scale-1}, is obtained. But here, differently from the isotropic case above for $m=0$, a nonzero dressed mass is obtained as
\begin{eqnarray}
\bar{m}^{2} &=& m^{2}\, \frac{\big\langle \hat{H}_{o}^{-1/2}\hat{a}^{6}(\phi)\hat{H}_{o}^{-1/2}\big\rangle \big\langle \hat{H}_{o}^{-1}\big\rangle^{\frac{1}{2}}}{\big\langle \hat{H}_{o}^{-1/2}\hat{a}^{4}(\phi)\hat{H}_{o}^{-1/2}\big\rangle ^{\frac{3}{2}}}=\frac{\tilde{m}^2}{\lambda^2}.\label{dressed-mass0}
\end{eqnarray}
This solution represents an isotropic dressed  spacetime, with the scale factor $\bar{a}(\phi)$, over which a massive mode with the mass $\bar{m}$, and an undressed wave-vector $\mathbf{k}=(k_{1},k_{2},k_{3})$
propagates.

\section{QFT on the dressed  spacetime \label{fluctuation}}

So far we have seen that the massive field modes propagating on an isotropic quantum geometry can explore a classical, anisotropic dressed background, $\tilde{g}_{ab}$ (see e.g., solutions (\ref{anisotropyBG})), which is the solution of Eqs.~(\ref{eq:main-sys}).
In this section, we will study the QFT and the issue of the gravitational particle productions on such an emergent anisotropic spacetime. 
%We will thus assume that, all modes of the test field, $\varphi(\phi, \mathbf{x})$,  probe the same anisotropic smooth background, with components given in (\ref{anisotropyBG}), so that an integration over all modes can be possible. 

\subsection{Field equation on the dressed anisotropic background\label{Field1}}

We suppose that the scalar field, $\varphi$, propagates on the dressed Bianchi I background
\begin{equation}
\tilde{g}_{ab}dx^{a}dx^{b}=-\tilde{N}_{\phi}^{2}\left(\phi\right)d\phi^{2}+\sum_{i=1}^{3}\tilde{a}_{i}^{2}\left(\phi\right)\left(dx^{i}\right)^{2},\label{BI-metric-class2}
\end{equation}
whose components are given as the solutions to  Eqs.~\eqref{eq:main-sys}. 
For conveniences, we introduce the new variables 
\begin{equation}
\tilde{c}(\phi)\coloneqq(\tilde{a}_{1}\tilde{a}_{2}\tilde{a}_{3})^{\frac{2}{3}}=(\tilde{c}_{1}\tilde{c}_{2}\tilde{c}_{3})^{\frac{1}{3}} \quad {\rm and} \quad \tilde{c}_{i}(\phi)\coloneqq\tilde{a}_{i}^{2}(\phi),
\end{equation} 
and consider a conformal time parameter, $\tilde{\eta}$, defined by
\begin{equation}
\tilde{N}_{\phi}d\phi=\tilde{N}_{\tilde{\eta}}d\tilde{\eta}=\tilde{c}^{1/2}d\tilde{\eta} \quad \Rightarrow \quad d\tilde{\eta}=\ell^{3}\langle\hat{H}_{o}^{-1}\rangle\tilde{c}(\phi)d\phi.
\label{conformaltime}
\end{equation}
%%%
Let's now take an auxiliary field, $\chi_{\mathbf{k}}\equiv\sqrt{\tilde{c}(\tilde{\eta})}\, \varphi_{\mathbf{k}}$.
Then, in terms of the above variables, the equation of motion  reads
\begin{equation}
\chi_{\mathbf{k}}^{\prime\prime}+\left[\tilde{\omega}_{\tilde{\eta},k}^{2}(\tilde{\eta})-\frac{\tilde{c}^{\prime\prime}}{2\tilde{c}}+\frac{\tilde{c}^{\prime2}}{4\tilde{c}^{2}}\right]\chi_{\mathbf{k}}=0,
\label{field-eq-mode1-2a}
\end{equation}
where, a prime stands for a differentiation with respect to
$\tilde{\eta}$, and  $\tilde{\omega}_{\tilde{\eta},k}$ is given by
\begin{equation}
\tilde{\omega}_{\tilde{\eta},k}^{2}(\tilde{\eta})=\tilde{c}^{-2}\left[k^{2}\frac{\big\langle \hat{H}_{o}^{-\frac{1}{2}}\hat{a}^{4}\hat{H}_{o}^{-\frac{1}{2}}\big\rangle }{\langle\hat{H}_{o}^{-1}\rangle}+m^{2}\frac{\big\langle \hat{H}_{o}^{-\frac{1}{2}}\hat{a}^{6}\hat{H}_{o}^{-\frac{1}{2}}\big\rangle }{\langle\hat{H}_{o}^{-1}\rangle}\right].
\label{omega-BI-1-b}
\end{equation}
Notice that, in simplifying equations above,  we have used the relations between components according to Eqs.~\eqref{eq:main-sys}. 

The frequency (\ref{omega-BI-1-b}) is specified  only when the solutions for $\tilde{c}(\tilde{\eta})$ are determined.
It turns out that, only
two classes of solutions for $\tilde{c}$ exist, which depend on the conditions on the field mass: 
\begin{enumerate}[label=\roman*)]
\item For massive field with {\em undressed} mass, $\tilde{m}=m\neq0$, the relation for
$\tilde{c}(\tilde{\eta})$ has the form
\begin{equation}
\tilde{c}(\tilde{\eta})=\frac{\big\langle \hat{H}_{o}^{-\frac{1}{2}}\hat{a}^{6}\hat{H}_{o}^{-\frac{1}{2}}\big\rangle ^{\frac{1}{3}}}{\langle\hat{H}_{o}^{-1}\rangle^{\frac{1}{3}}}.\label{ctilde-1}
\end{equation}
%%%%%%%%%%%%%%
\item For a massive field with the {\em dressed} mass, $\tilde{m}\neq m$, or a {\em massless} field, $\tilde{c}(\tilde{\eta})$ has the solutions of the form 
\begin{equation}
\tilde{c}(\tilde{\eta})=\xi^{2}\frac{\big\langle \hat{H}_{o}^{-\frac{1}{2}}\hat{a}^{4}\hat{H}_{o}^{-\frac{1}{2}}\big\rangle ^{\frac{1}{2}}}{\langle\hat{H}_{o}^{-1}\rangle^{\frac{1}{2}}}=\xi^{2}\tilde{a}^{2}(\tilde{\eta}),\label{ctilde-2}
\end{equation}
where,   $\xi$ is a  parameter which distinguishes the anisotropic solutions from the isotropic one, and depends on the conditions imposed to the additional unknown variables $\tilde{a}_i$'s and $\tilde{k}_i$'s of the system (\ref{eq:main-sys}). The case $\xi^{2}=1$  associates with the {\em isotropic} solution,
whereas the case $\xi\neq 1$ denotes the {\em anisotropic} solutions  [cf. Eqs.~(\ref{massless-dressed-scale}) and (\ref{dressed-mass0}) for  different classes of the  solutions].
\end{enumerate}

\subsection{The adiabatic condition and the vacuum state\label{Field2}}

Any solution $\chi_{\mathbf{k}}(\tilde{\eta})$ to the Klein-Gordon (\ref{field-eq-mode1-2a})
can be expanded as 
\begin{equation}
\chi_{\mathbf{k}}(\tilde{\eta})=\frac{1}{\sqrt{2}}\left[a_{\mathbf{k}}u_{k}^{\ast}(\tilde{\eta})+a_{-\mathbf{k}}^{\ast}u_{k}(\tilde{\eta})\right],
\label{mode-decompose1}
\end{equation}
where, $a_{\mathbf{k}}$ and $a_{\mathbf{k}}^{\ast}$ are constants
of integration, and $u_{k}(\tilde{\eta})$ are the general mode solutions of Eq.~(\ref{field-eq-mode1-2a}), 
\begin{equation}
u_{k}^{\prime\prime}+\Big(\tilde{\omega}_{\tilde{\eta},k}^{2}(\tilde{\eta})-{\cal Q}(\tilde{\eta})\Big)u_{k}=0,
\label{KG-1-eff-2}
\end{equation}
with, 
${\cal Q}\equiv\tilde{c}^{\prime\prime}/2\tilde{c}
-\tilde{c}^{\prime2}/4\tilde{c}^{2}$. These modes, together with their complex conjugates,
$u_{k}^{\ast}(\tilde{\eta})$, form a complete set of orthonormal basis, under the scalar product 
\begin{equation}
W(u_{k}^{\ast},u_{k})\coloneqq u_{k}u_{k}^{\ast\prime}-u_{k}^{\ast}u_{k}^{\prime}=2i.\label{Wronskian}
\end{equation}

In quantum theory, $a_{\mathbf{k}}$ and $a_{\mathbf{k}}^{\ast}$ become annihilation and creation operators, $\hat{a}_{\mathbf{k}}$ and $\hat{a}_{\mathbf{k}}^{\dagger}$,
 satisfying the commutation relation 
\begin{equation}
\big[\hat{a}_{\mathbf{k}},\hat{a}_{\mathbf{k}^{\prime}}^{\dagger}\big]=\hbar\ell^{3}\delta_{\mathbf{k},\mathbf{k}^{\prime}},
\end{equation}
where, all other commutation relations vanish. For the  \emph{positive frequency} solutions, a choice of the basis, $u_{k}(\tilde{\eta})$, determines
a vacuum state, $|0\rangle$, which is defined as the eigenstate
of the annihilation operators obeying the equation $\hat{a}_{\mathbf{k}}|0\rangle=0$. Different families of the mode solutions distinguish different vacuum states. 
Accordingly,  a Fock space is generated by
repeatedly acting the operator $\hat{a}_{\mathbf{k}}^{\dagger}$ on a chosen  vacuum state.
Thereby,  the evolution of the quantum fields on the quantum-gravity-induced dressed spacetime is determined.
Now, the general (auxiliary) field solutions, $\chi(\tilde{\eta},\mathbf{x})$, can be expanded as
\begin{equation}
\chi(\tilde{\eta},\mathbf{x})=\frac{1}{\ell^{3}}\sum_\mathbf{k} \chi_{\mathbf{k}}(\tilde{\eta},\mathbf{x}),\label{mode-decompose2}
\end{equation}
where,
\begin{equation}
\chi_{\mathbf{k}}=\frac{1}{\sqrt{2}}\left[a_{\mathbf{k}}u_{k}^{*}(\tilde{\eta})e^{i\mathbf{k}\cdot\mathbf{x}}+a_{\mathbf{k}}^{\ast}u_{k}(\tilde{\eta})e^{-i\mathbf{k}\cdot\mathbf{x}}\right].\label{mode-decompose2b}
\end{equation}

To have a well-defined vacuum state for the mode solutions, the (effective) frequency
\begin{equation}
\Omega_{k}^{2}(\tilde{\eta}) = \tilde{\omega}_{\tilde{\eta},k}^{2}(\tilde{\eta})-{\cal Q}(\tilde{\eta}),
\end{equation}
should be positive. Thus, the wave-number must satisfy
$k^{2}\geq k_{\ast}^{2}$, where 
\begin{equation}
k_{\ast}^{2}(\tilde{\eta})\coloneqq
\frac{{\cal Q}(\tilde{\eta})\tilde{c}^{2}(\tilde{\eta})\langle\hat{H}_{o}^{-1}\rangle}{\big\langle \hat{H}_{o}^{-\frac{1}{2}}\hat{a}^{4}\hat{H}_{o}^{-\frac{1}{2}}\big\rangle}-m^{2}\, \frac{\big\langle \hat{H}_{o}^{-\frac{1}{2}}\hat{a}^{6}\hat{H}_{o}^{-\frac{1}{2}}\big\rangle}{\big\langle \hat{H}_{o}^{-\frac{1}{2}}\hat{a}^{4}\hat{H}_{o}^{-\frac{1}{2}}\big\rangle},
\label{k-star}
\end{equation}
introduces a physical wavelength, $\lambda_\ast(\tilde{\eta})\equiv2\pi\sqrt{\tilde{c}(\tilde{\eta})}/k_{\ast}$, which provides an upper bound for the wavelengths of the physical mode functions, i.e.,  $\lambda\leq\lambda_\ast$. Therefore,  modes with short wavelengths (i.e., large momenta), $\lambda\ll\lambda_\ast$, characterize
the vacuum states in short distances, i.e., the \emph{ultra-violet} (UV) regimes. This regime contains a limit of arbitrary slow time, $\tilde{\eta}$, variation
of the metric functions, called the \emph{adiabatic} regime.
We will, henceforth, explore the mode functions describing the physical vacuum and particles, associated with this adiabatic regime.

A positive-frequency, adiabatic vacuum mode can be defined by a generalized WKB approximate solution to the Klein-Gordon equation,  as \cite{Parker:1974qw}
\begin{equation}
\underline{u}_{k}(\tilde{\eta})=\frac{1}{\sqrt{W_{k}(\tilde{\eta})}}\exp\Big(-i\int^{\tilde{\eta}}W_{k}(\eta)d\eta\Big).
\label{adibatic-1}
\end{equation}
Substituting this relation in Eq.~(\ref{KG-1-eff-2}) we obtain a relation for $W_k(\eta)$ as
\begin{equation}
\frac{W_k^{\prime\prime}}{W_k}-\frac{W^{\prime}_k}{W_k^2}-\frac{1}{2}\frac{W^{\prime 2}_k}{W_k^2}+2\left(W_k^2-\Omega_{k}^{2}\right)=0.
\label{KG-1-eff-2W}
\end{equation}
An appropriate
function  can be chosen for $W_{k}(\tilde{\eta})$, such as  \cite{Chakraborty:1973}
\begin{equation}
W_{k}(\tilde{\eta})=\big[Y(1+\underline{\epsilon_{2}})(1+\underline{\epsilon_{4}})\big]^{\frac{1}{2}},\label{W-k-1}
\end{equation}
where, $Y\equiv\Omega_{k}^{2}=\tilde{\omega}_{\tilde{\eta},k}^{2}-{\cal Q}$,
and 
\begin{align}
\underline{\epsilon_{2}} & \coloneqq-Y^{-\frac{3}{4}}\partial_{\tilde{\eta}}\Big(Y^{-\frac{1}{2}}\partial_{\tilde{\eta}}Y^{\frac{1}{4}}\Big),\label{epsilon-under-1}\\
\underline{\epsilon_{4}} & \coloneqq-Y^{-\frac{1}{2}}(1+\underline{\epsilon_{2}})^{-\frac{3}{4}}\partial_{\tilde{\eta}}\Big\{[Y(1+\underline{\epsilon_{2}})]^{-\frac{1}{2}}\partial_{\tilde{\eta}}(1+\underline{\epsilon_{2}})^{\frac{1}{4}}\Big\}.\label{epsilon-under-2}
\end{align}
To get the solutions for $\underline{u}_{k}(\tilde{\eta})$, one needs to solve the equation (\ref{KG-1-eff-2W}) for $W(\tilde{\eta})$.
However, instead of solving (\ref{KG-1-eff-2W}) exactly, one way is to generate
asymptotic series in orders of time $\tilde{\eta}$ derivatives
of the background dressed metric. Terminating this series at a given order
will specify an adiabatic mode $\underline{u}_{k}(\tilde{\eta})$ to
that order. 
Up to an order $n$ (being the power of $\tilde{\eta}$-derivatives of  $\tilde{a}_{i}$), denoted ${\cal O}(\sqrt{\tilde{c}}/k\lambda_{n+\varepsilon})^{n+\varepsilon}$ (with positive real number $\varepsilon$),
 the asymptotic adiabatic expansion of 
$W_{k}$ is defined to match
\begin{equation}
W_{k}(\tilde{\eta})=W_{k}^{(0)}+W_{k}^{(2)}+\cdot\cdot\cdot+W_{k}^{(n)},
\label{W-expansion}
\end{equation}
and, is obtained for higher order estimates by iteration. 

When computing the expectation value of the energy-momentum operator of the scalar field, $\langle\tilde{0}|\hat{T}_{ab}|\tilde{0}\rangle$,
with respect to the adiabatic vacuum, $|\tilde{0}\rangle$, associated to the mode function (\ref{adibatic-1}), one encounters the UV divergences issues. All of these UV divergences are contained within
the terms of adiabatic order equal to and smaller than four
[cf. \cite{Fulling:1974pu} and the appendix of Ref.~\cite{Tavakoli:2015fvz}]. Thus, the (adiabatic) regularization of the energy density and pressure of the scalar field are obtained from  subtractions of the divergences contained within  the adiabatic  terms up to fourth order; e.g., the renormalized energy density of the field is given by
\begin{equation}
\langle\tilde{0}|\hat{\rho}_{\varphi}|\tilde{0}\rangle_{\rm ren} = \frac{\hbar}{\ell^3\tilde{c}^2} \sum_{k} \left(\rho_{k}[u_k(\tilde{\eta})]- \underline{\rho_{k}}[\underline{u_k}(\tilde{\eta})]\right),
\end{equation}
where, $\rho_k$ is the energy density of each mode, $u_k(\tilde{\eta})$, and $\underline{\rho_{k}}$ is the energy density associated with the mode, $\underline{u_k}$, up to 4th  adiabatic order.
We will, thus, restrict
ourselves to the fourth order adiabatic states only\footnote{Note that, our aim at the moment is just to avoid the UV divergence terms when regularizing the energy-momentum of the field. However,  to have a more complete mode solution, in particular when exploring the issues of the particle productions or backreaction effects, as we will see later, higher order terms in $W_k(\tilde{\eta})$ should be taken into account.}. Then, by setting $n=4$,
we define $W_{k}(\tilde{\eta})$ to match the terms in (\ref{W-expansion})
that fall slowly in $\tilde{\omega}_{\tilde{\eta},k}$, rather than
considering the exact
mode solutions of (\ref{KG-1-eff-2}). 
Accordingly, we consider an appropriate asymptotic condition (up to the order four) at which  the field's exact mode functions, $u_{k}(\tilde{\eta})$, match the adiabatic functions, $\underline{u}_{k}(\tilde{\eta})$, as
\begin{subequations}
	\label{adiab-cond2}
	\begin{align}
	|u_{k}(\tilde{\eta}_{b})|= & \big|\underline{u}_{k}(\tilde{\eta}_{b})\big|\Big(1+{\cal O}\big(\sqrt{\tilde{c}}/k\lambda_{4+\varepsilon}\big)^{4+\varepsilon}\Big), \\
	|u_{k}^{\prime}(\tilde{\eta}_{b})|= & \big|\underline{u}_{k}^{\prime}(\tilde{\eta}_{b})\big|\Big(1+{\cal O}\big(\sqrt{\tilde{c}}/k\lambda_{4+\varepsilon}\big)^{4+\varepsilon}\Big),
	\end{align}
\end{subequations}
where, $\tilde{\eta}=\tilde{\eta}_{b}$ is a natural choice for the preferred instant of time, given at the quantum bounce, in LQC.
When a mode function,  $u_{k}(\tilde{\eta})$, satisfies the conditions (\ref{adiab-cond2}), at some initial time $\tilde{\eta}_{b}$, it will satisfy it for all times $\tilde{\eta}$. Therefore, an observable
vacuum, $|0\rangle$, associated to $u_{k}(\tilde{\eta})$,  will be of the 4th order, for all times.

Following the above discussion, we thus discard the adiabatic order terms higher than four in Eq.~(\ref{W-k-1}), and write
\begin{equation}
W_{k}(\tilde{\eta})=\tilde{\omega}_{\tilde{\eta},k}(1+\epsilon_{2}+\epsilon_{4})^{\frac{1}{2}},\label{adibatic-2}
\end{equation}
where,  $\epsilon_{2}$, $\epsilon_{4}$ are defined by
\begin{align}
\epsilon_{2} & = \underline{\epsilon_{2}}-\tilde{\omega}_{\tilde{\eta},k}^{-2}\, {\cal Q}\nonumber\\
& = -\frac{1}{4}Y^{-2}Y^{\prime\prime}+\frac{5}{16}Y^{-3}(Y^{\prime})^{2}
-\tilde{\omega}_{\tilde{\eta},k}^{-2}\, {\cal Q}, \label{adibatic-2-sec1} 
\\
\epsilon_{4} & =\underline{\epsilon_{4}}-\underline{\epsilon_{2}}\, \tilde{\omega}_{\tilde{\eta},k}^{-2}\, {\cal Q} \nonumber \\
&= -\frac{1}{4}Y^{-1}(1+\underline{\epsilon_{2}})^{-2}\Big[\underline{\epsilon_{2}}^{\prime\prime}-\frac{1}{2}Y^{-1}Y^{\prime}\underline{\epsilon_{2}}^{\prime}-\frac{5}{4}(1+\underline{\epsilon_{2}})^{-1}(\underline{\epsilon_{2}}^{\prime})^{2}\Big]-\underline{\epsilon_{2}}\, \tilde{\omega}_{\tilde{\eta},k}^{-2}\, {\cal Q},
\label{adibatic-2-sec}
\end{align}
and,   $\underline{\epsilon_{2}}$ is given by
\begin{equation}
\underline{\epsilon_{2}}=-\frac{1}{4}Y^{-2}Y^{\prime\prime}+\frac{5}{16}Y^{-3}(Y^{\prime})^{2}.%
\label{eps-1}
\end{equation}
The leading order term of $\underline{\epsilon_{2}}$ is of second order, whereas the leading order in $\underline{\epsilon_{4}}$ is
four. Thus, $\epsilon_{2}$ contains terms of orders two, four and
higher, i.e, 
\begin{equation}
\epsilon_{2}=\epsilon_{2}^{(2)}+\epsilon_{2}^{(4)}+{\rm Higher~order~terms},\label{adibatic-2-sec1App-1}
\end{equation}
and, $\epsilon_{4}$ contains leading order term four, i.e., 
\begin{align}
\epsilon_{4} &=  \epsilon_{4}^{(4)}~+~{\rm Higher~order~terms}. \label{adibatic-2-sec-App-2}
\end{align}
It should be noticed that, we have considered only terms until the fourth order in their expressions [for more details, see \cite{Tavakoli:2015fvz}].

%%%%%%%%%%%%%%%%%%%%%%%%%%%%%
\section{Gravitational particle production}

Our aim in this section is to study the gravitational particle production in the herein quantum gravity regime  due to quantum field on the  anisotropic, dressed background.

%\subsection{Particle productions in the adiabatic regime}

Once a vacuum state, $|\tilde{0}\rangle$, is specified due to the positive frequency solutions, $\underline{v}_{k}(\tilde{\eta})$, of Eq.~(\ref{KG-1-eff-2}), a 
Fock space, ${\cal H}_{{\rm F}}$,  for the quantum field $\varphi$ is generated. Let $\{v_{k}\}$ and $\{\underline{v}_{k}\}$ be two sets of WKB solutions, given by Eq.~(\ref{adibatic-1}), in the herein adiabatic regime. These mode functions form two {\em normalized} bases\footnote{They satisfy the normalization condition (\ref{Wronskian}) for the mode functions.} for  ${\cal H}_{{\rm F}}$, so  they can be related to each other through the time-independent Bogolyubov coefficients, $\alpha_{k}$ and $\beta_{k}$, as
\begin{equation}
\underline{v}_{k}(\tilde{\eta})=\alpha_{k}v_{k}(\tilde{\eta})+\beta_{k}v_{k}^{\ast}(\tilde{\eta}).\label{Bog-relation1}
\end{equation}
The Bogolyubov coefficients 
satisfy the relation $|\alpha_{k}|^{2}-|\beta_{k}|^{2}=1$ via condition (\ref{Wronskian}). 
Comparing   
Eqs.~(\ref{Bog-relation1}) and (\ref{mode-decompose2b}), it follows that
the creation and annihilation operators
associated with  two families of mode functions (i.e, those with and without `underline'), are related as
\begin{equation}
\hat{a}_{\mathbf{k}}=\alpha_{k}\underline{\hat{a}}_{\mathbf{k}}+\beta_{k}^{\ast}\underline{\hat{a}}_{\mathbf{k}}^{\dagger}.
\end{equation}
Working in Heisenberg picture, an initial vacuum state of the system, say $|\tilde{0}\rangle$ connected to the `underlined' modes, $\underline{v}_{k}(\tilde{\eta})$,  is the vacuum state of the system for all times. Then, the  number operator, $\hat{N}_{\mathbf{k}}=(\hbar\ell^{3})^{-1}\hat{a}_{\mathbf{k}}^{\dagger}\hat{a}_{\mathbf{k}}$,
associated to the particles in $v_{k}$ mode, gives  the
average number of particles in the $|\tilde{0}\rangle$ vacuum. 
Thus, the $v_{k}$-mode related vacuum state  contains
\begin{equation}
{\cal N}_{k}\coloneqq\langle\tilde{0}|\hat{N}_{\mathbf{k}}|\tilde{0}\rangle=|\beta_{k}|^{2},
\label{NumberParticle1}
\end{equation}
particles in the $\underline{v}_{k}$-mode vacuum.

Let us rewrite the mode solution, $\underline{v}_{k}$,  in the WKB
approximation (\ref{adibatic-1}), as 
\begin{equation}
\underline{v}_{k}(\tilde{\eta})=\frac{1}{\sqrt{W_{k}(\tilde{\eta})}}\Big[\alpha_{k}e_{k}(\tilde{\eta})+\beta_{k}e_{k}^{\ast}(\tilde{\eta})\Big],\label{beta-particle-1}
\end{equation}
where,
\begin{equation}
e_{k}(\tilde{\eta})\coloneqq\exp\Big(-i\int^{\tilde{\eta}} d\eta\,  W_k(\eta)\Big).
\end{equation}
Taking the time
($\tilde{\eta}$) derivative of $\underline{v}_{k}(\tilde{\eta})$, we obtain 
\begin{equation}
\underline{v}_{k}^{\prime}(\tilde{\eta})=-i\sqrt{W_{k}}\Big[\alpha_{k}e_{k}(\tilde{\eta})-\beta_{k}e_{k}^{\ast}(\tilde{\eta})\Big].\label{beta-particle-2}
\end{equation}
Inverting the Eqs.~(\ref{beta-particle-1})
and (\ref{beta-particle-2}) yields a relation for $\beta_k$ as
\begin{equation}
\beta_{k}=\frac{\sqrt{W_{k}}}{2}\left(\underline{v}_{k}-\frac{i}{W_{k}}\underline{v}_{k}^{\prime}\right)e_{k}.\label{beta_rel1}
\end{equation}
Since the initial condition in LQC is fixed at the quantum bounce,  $\tilde{\eta}=\tilde{\eta}_{b}$, by 
setting  $\alpha_{k}=1$ and $\beta_{k}=0$ at
$\tilde{\eta}=\tilde{\eta}_{b}$, we assume that no particle is created at bounce.
This yields the following conditions on $\underline{v}_{k}$:
\begin{equation}
\underline{v}_{k}(\tilde{\eta}_{b})=1/W_{k}(\tilde{\eta}_{b}) \quad \text{and} \quad \underline{v}_{k}^{\prime}(\tilde{\eta}_{b})=-iW_{k}(\tilde{\eta}_{b})\, \underline{v}_{k}(\tilde{\eta}_{b}),
\end{equation} 
where $e_{k}(\tilde{\eta}_{b})=1$. Then, for any time $\tilde{\eta}>\tilde{\eta}_b$, the number of particle
production becomes
\begin{equation}
{\cal N}_{k}(\tilde{\eta})=\frac{1}{4}\Big(W_{k}\, |v_{k}|^{2}+W_{k}^{-1}|v_{k}^{\prime}|^{2}-2\Big).\label{Part-number}
\end{equation}
Here, for conveniences, we have dropped the `underline' for the  mode functions after the bounce ($\tilde{\eta}>\tilde{\eta}_b$), i.e., we assume
$\underline{v}_k\equiv \underline{v}_{k}(\tilde{\eta}_b)$  and $v_k(\tilde{\eta})\equiv \underline{v}_{k}(\tilde{\eta})$ for $\tilde{\eta}>\tilde{\eta}_b$.

Having the number of particle production per mode,  ${\cal N}_{k}(\tilde{\eta})$, at a given time $\tilde{\eta}>\tilde{\eta}_b$, we can compute the total number density of production, ${\cal N}(\tilde{\eta})$, as the limit of $\sum_{k}{\cal N}_k(\tilde{\eta})$ in a box of volume $\ell^3\to \infty$, divided by the volume of the universe, $V=\ell^3|\tilde{a}_1\tilde{a}_2\tilde{a}_3|=\ell^3\tilde{c}^{3/2}$, as
\begin{equation}
{\cal N}(\tilde{\eta}) = \frac{\ell^3}{V} \sum_{k}{\cal N}_k(\tilde{\eta}) = \frac{1}{4\tilde{c}^{3/2}} \sum_k \Big(W_{k}\, |v_{k}|^{2}+W_{k}^{-1}|v_{k}^{\prime}|^{2}-2\Big).
\end{equation}
The  energy density of the created particles, for each mode, reads $\varrho_{k}(\tilde{\eta})\equiv W_{k}(\tilde{\eta})\, {\cal N}_{k}(\tilde{\eta})$.
Thus, the total energy density is obtained by summing over all  modes, as 
\begin{equation}
\rho_{{\rm par}}=\frac{1}{\tilde{c}^2} \sum_{k}\, 
\varrho_{k}(\tilde{\eta}),\label{energy-part-tot}
\end{equation}
where,
\begin{align}
\varrho_{k}(\tilde{\eta}) =  \frac{1}{4}\Big(|v_{k}^{\prime}|^{2}+W_{k}^2\, |v_{k}|^{2}-2W_{k}\Big).\label{energy-part-tot1}
\end{align}
At the   bounce, $\tilde{\eta}=\tilde{\eta}_{b}$,
the energy density of production is zero, $\varrho_{k}(\tilde{\eta}_{b})=0$, as expected.
However, for $\tilde{\eta}>\tilde{\eta}_{b}$, particles will be produced as the universe expands. In the following, we will analyze the energy density of production in the  assumed adiabatic regime.

Following the (adiabatic) regularization scheme, the energy density  of created particles  is obtained as
\begin{equation}
\rho_{\rm par}^{\rm (ren)}=\frac{1}{\tilde{c}^{2}}\sum_{k}\Big(\varrho_{k}[v_{k}(\tilde{\eta})]-\underline{\varrho_{k}}(\tilde{\eta})\Big),
\label{particle-ren}
\end{equation}
where,
\begin{align}
\underline{\varrho_{k}}(\tilde{\eta}) &=   \frac{1}{16}\left[\frac{(\tilde{\omega}_{k,\tilde{\eta}}^{\prime})^{2}}{\tilde{\omega}_{k,\tilde{\eta}}^{3}}+\tilde{\omega}_{k,\tilde{\eta}}\big(\epsilon_{2}^{(2)}\big)^{2}-\frac{(\tilde{\omega}_{k,\tilde{\eta}}^{\prime})^{2}}{2\tilde{\omega}_{k,\tilde{\eta}}^{3}}\epsilon_{2}^{(2)}+\frac{\tilde{\omega}_{k,\tilde{\eta}}^{\prime}}{\tilde{\omega}_{k,\tilde{\eta}}^{2}}\epsilon_{2}^{\prime(3)}\right]\nonumber \\
& =  \varrho_{k}^{(0)}+\varrho_{k}^{(2)}+\varrho_{k}^{(4)}.
\label{energy-reg}
\end{align}
It turns out that, $\underline{\varrho_{k}}$ does not fall off faster than $k^{-4}$
when $k\rightarrow\infty$.
The zeroth adiabatic order
term in (\ref{energy-reg}) is zero, $\varrho_{k}^{(0)}=0$, and the divergences
are included in the second and fourth order terms (for massive modes).
For the massless modes, the divergences are included only in the
fourth order term. 
Therefore, the renormalized energy density of
created particles, Eq.~(\ref{particle-ren}), can be obtained by subtraction of the adiabatic
vacuum energy of the particle productions up to fourth order. When considering the higher order terms (more than 4th order) in the adiabatic mode functions, $v_k(\tilde{\eta})$, associated to the vacuum state $|0\rangle$,  particles are produced and the total energy density of the created particles is  proportional to $1/\tilde{c}^2(\tilde{\eta})$. For an isotropic case (either massive or massless), where $\tilde{c}=\tilde{a}^2$, we get $\rho_{\rm par}^{\rm (ren)}\propto 1/\tilde{a}^4(\tilde{\eta})$ which scales as radiation. This is a similar situation derived in Ref.~\cite{Graef:2020qwe}. Therefore, unlike the classical FLRW cosmology, even massless modes contain nonzero particle production due to  quantum gravity effects.

In the pre-inflationary scenario considered in \cite{Graef:2020qwe}, by assuming a massless field in an isotropic quantum cosmological background, $\tilde{a}(\tilde{\eta})=\tilde{c}^{1/2}(\tilde{\eta})$, 
an integration range were chosen in Eq.~(\ref{energy-part-tot}) by the window of observable modes of the CMB.
Therein, it has been argued that, when
taking a UV cutoff at the characteristic momentum $k_b=k_\ast(\tilde{\eta}_b)=\sqrt{\tilde{a}^{\prime\prime}(\tilde{\eta})/\tilde{a}(\tilde{\eta})}|_{\tilde{\eta}_b}\approx3.21\, m_{\rm Pl}$, and an {\em infra-red} (IR) cutoff at $k_{\ast}(\tilde{\eta})=\sqrt{\tilde{a}^{\prime\prime}(\tilde{\eta})/\tilde{a}(\tilde{\eta})}=\lambda^{-1}_{\ast}$ for all $\tilde{\eta}>\tilde{\eta}_b$, being the physical energy of particles after they reenter the effective horizon, the energy density of created particles becomes
\begin{equation}
\rho_{\rm par} = 0.012\, m_{\rm Pl}^4/\tilde{a}^4(\tilde{\eta}).
\label{density-particle-rad}
\end{equation}
This gives the energy density of  particles produced in the region  $k_{\ast}\leq k \leq k_b$. 
Thus, the main contribution to the particle productions are the modes whose wavelengths, $\lambda$, hold the range $\lambda_b\leq\lambda\leq\lambda_{\ast}$, i.e., the modes that reenter $\lambda_{\ast}$ after the bounce, during the pre-inflationary phase, and will only reexit $\lambda_{\ast}$ again in the slow-roll inflationary phase.

This implies that the energy density of particles produced in the quantum gravity regime,  which scales as relativistic matter, is significant comparing to the background energy density. The background energy density during the slow-roll inflation is dominant over the backreaction of particle productions. However, it should be guaranteed that the production  density is still dominant  in the pre-inflationary phase before beginning of the slow-roll phase. 
By assuming that, the elapsed e-foldings between the bounce and the onset of inflation is about $4$-$5$ e-folds, the energy density of the backreaction is smaller than the background energy density as  $\rho_{\rm par}\lesssim \langle\hat{\rho}_{\varphi}\rangle\approx 2\times 10^{-5}\, m_{\rm Pl}^4$. However, the analysis in \cite{Graef:2020qwe} indicates that, the energy density (\ref{density-particle-rad})  is 2 orders of magnitude larger than the required upper bound, $2\times 10^{-5}~m_{\rm Pl}^4$, estimated for the density of particle creation during the pre-inflationary phase.
It turns out that, the backreaction of produced particles cannot be neglected, so that a more careful analysis of backreacted wave function for the background quantum geometry is needed. 

The above argument indicates that, when starting from Eq.~(\ref{const-tot1}), the backreaction of the field modes on the quantum background is not negligible, so that the total wave function, $\Psi(\nu, Q_\mathbf{k}, \phi)$, cannot be decomposed as $\Psi=\psi_o\otimes \psi$. It may even make a further constraint on expanding the right-hand-side of Eq.~(\ref{const-tot1}) to derive the evolution equation (\ref{shro-eq2-approx}). Nevertheless, if we suppose that the approximation (\ref{shro-eq2-approx}) is valid, the presence of  the backreaction would lead to a modification of the total state as $\Psi=\psi_o\otimes \psi + \delta\Psi$  [cf. \cite{Dapor:2012jg,Lewandowski:2017cvz}]. Taking into account such modification,   the dispersion relation of the field, propagating on the emergent  effective geometry, will be modified such that the local Lorentz symmetry becomes violated. In such a scenario, each mode feels a distinctive background geometry which depends on that mode; a rainbow dressed background emerges. On such backgrounds,  the standard approach for studying the infinite number of field modes  fails and an alternative procedure should be employed [e.g., see \cite{Martin:2000xs}]. For other approaches in canonical quantum gravity, where the quantum theory of cosmological perturbations and their backreactions are implemented, see e.g. \cite{Bojowald:2020emy,Fernandez-Mendez:2013jqa,Gomar:2014faa,Gomar:2015oea,Bojowald:2008gz,Cailleteau:2012fy,Cailleteau:2011kr}.

%%%%%%%%%%%%%%%%%%%%%%%%%
\section{Conclusion and Discussion \label{conclusion}}

In this paper, the quantum theory  of a (inhomogeneous) {\em massive}  test field, $\varphi$, propagating on a  quantized FLRW geometry is addressed. 
The background geometry constitutes of a homogeneous massless scalar field, $\phi$, as matter source, which plays the role of internal time in quantum theory.
From an effective point of view, due to quantum gravity effects on the background geometry, quantum modes of the field can experience a dressed  spacetime whose geometry differs from the original FLRW metric. 
If the backreaction of the field modes, $Q_{\mathbf{k}}$, on the background is discarded, within a {\em test field} approximation,  the full quantum state of the system can be decomposed as $\Psi(\nu,Q_{\mathbf{k}},\phi)=\Psi_{o}(\nu,\phi)\otimes\psi(Q_{\mathbf{k}},\phi)$;  $\Psi_{o}$ and $\psi$ are the  quantum states of the (unperturbed) background and  the
 field modes, respectively.

In the interaction picture, an evolution equation emerges for $\psi(Q_{\mathbf{k}},\phi)$  which resembles the Schr\"odinger equation for the same field modes propagating on a time-dependent dressed spacetime, whose metric components are functions of the quantum fluctuations of the FLRW geometry.
For massive and massless modes, there exists a wide class of solutions for the effective dressed background metric. 
The \emph{massless} modes can only experience an isotropic and homogeneous dressed background with a dressed scale factor,
$\tilde{a}(\phi)$ [cf. see  \cite{Ashtekar:2009mb}].
The \emph{massive} modes, however, yield a general class of solutions for the  emergent dressed geometries which resembles the anisotropic  Biachi I spacetimes.
Likewise, the scale factors, $(\tilde{a}_{1},\tilde{a}_{2},\tilde{a}_{3})$, of the consequent dressed Bianchi-I metric are functions of fluctuations of the isotropic quantum geometry.

Given a dressed anisotropic spacetime, as a solution discussed above, we reviewed the standard quantum field theory  on such   background.
More precisely, we investigated the issue of gravitational particle production associated to the  field modes on the dressed Bianchi I geometry in a suitably chosen adiabatic regime.
This led to some  backreaction issues in the super-Planckian regime which may affect the dynamics of the early universe. To have a regularized  energy-momentum operator of the test field, the adiabatic vacuum state was chosen up to 4th order terms. We computed the energy density of the particle production within this adiabatic limit. The divergences in the energy density of the produced particles were regularized within the fourth order adiabatic terms and the remaining terms change as $\rho_{\rm par}^{\rm (ren)}\propto  1/(\tilde{a}_1\tilde{a}_2\tilde{a}_3)^{4/3}$. Some phenomenological issues related to such particle production were discussed. It was demonstrated that the backreaction due to the particle production in the super-Planckina regime may have significant effects on the evolution of the universe and may subsequently  modify the existing pre-inflationary scenario of LQC \cite{Agullo:2012sh} [cf. \cite{Graef:2020qwe}]

\begin{acknowledgments}
The results of this review article are mainly based on the past collaborations with my colleagues, J. C. Fabris and S. Rastgoo. This paper
is based upon work from European Cooperation in Science and Technology
(COST) action CA18108 -- Quantum gravity phenomenology in the multi-messenger
approach supported by COST. 
\end{acknowledgments}

\bibliography{Bibliography}

\end{document}